\documentclass[preprintnumbers,twocolumn,aps,floatfix]{revtex4}
\usepackage{psfig}
\begin{document}
\preprint{UH-511-1057-2004}
\preprint{LBL-56337}
\title{A prediction for $\left|{\cal U}_{e3}\right|$ from patterns in the charged lepton spectra}
\author{Javier Ferrandis $^{1,2}$} 
\email{ferrandis @ mac.com}
\homepage{http: // homepage.mac.com / ferrandis}
\author{Sandip Pakvasa $^{1}$} 
\email{pakvasa @ phys.hawaii.edu}
\affiliation{ $^{1}$ Department of Physics \& Astronomy \\
 University of Hawaii at Manoa \\
 2505 Correa Road \\
 Honolulu, HI, 96822\\}
 \affiliation{ $^{2}$ Theoretical Physics Group \\
 Lawrence Berkeley National Laboratory  \\
One Cyclotron Road, Berkeley CA 94720}
\begin{abstract}{
 It is shown that empirical relations between 
the charged lepton spectra and the quark spectra together with
a bimaximal or near bimaximal neutrino mixing matrix 
necessarily imply that there is a contribution to $\left|{\cal U}_{e3}\right|$ 
given by $\theta_{C}/3\sqrt{2} \approx \sqrt{m_e/2m_\mu} \approx 0.052$, where $\theta_{C}$ is the Cabibbo angle.
This prediction could be tested in the
near future reactor experiments. The charged lepton mixing also generates
a less robust prediction for the angle $\theta_{23}$ and a small contribution to the phase $\delta$.}
\end{abstract}
\maketitle
\newpage
%
\section{Introduction}
During the last year
our knowledge of the leptonic mixing matrix
has reached the precision level. 
The most recent $90$\% C.L. experimental 
results \cite{Ashie:2004mr,Araki:2004mb,Apollonio:2002gd}
and several global fits \cite{Bahcall:2004ut,Gonzalez-Garcia:2004it,Bandyopadhyay:2004da,Maltoni:2004ei}
have improved our knowledge of the neutrino mass differences and
indicate that the atmospheric mixing is almost maximal while the
the solar mixing deviates from maximality in a particular way. In the standard notation,
\begin{eqnarray}
\sin \theta_{12} &=& 0.53 \pm 0.04   , \\
\sin \theta_{23} &=& 0.70\pm 0.11   , \\
\sin \theta_{13} &<& 0.15  ,\\
\Delta m^{2}_{\rm sun} &=& \Delta m^{2}_{\rm 21}  = (8.2 \pm 0.6)  \times 10^{-5} {\rm eV}^{2} , \\
\Delta m^{2}_{\rm atm}& =&  \Delta m^{2}_{\rm 23}  = (2.45 \pm0.55)\times 10^{-3}{\rm eV}^{2} , 
\end{eqnarray}
This substantial improvement has confirmed that the leptonic mixing matrix, 
called MNSP matrix,  is nearly bimaximal \cite{bimaximal,morebimax} and the particular 
deviation from bimaximality observed has revealed a surprising relation 
between the Cabibbo angle, $\theta_{C}$ 
and the solar mixing angle  \cite{Rodejohann:2003sc}
sometimes called the 
quark-lepton complementarity relation, 
$\theta_{C} + \theta_{12} \approx \pi/4$, {\it hereafter} referred to as
QLC relation. 
Therefore based on the experimental 
data it is convenient to define the following parametrization
\cite{Rodejohann1} of the MNSP matrix, 
\begin{equation}
{\cal V}_{\rm MNSP}=
\left[
\begin{array}{ccc}
 \frac{1}{\sqrt{2}} \left( 1 + \lambda \right)  & - \frac{1}{\sqrt{2}} \left( 1 - \lambda \right) & 0 \\
 \frac{1}{2} \left( 1 - \lambda \right) &  \frac{1}{2} \left( 1 + \lambda \right) &  -\frac{1}{\sqrt{2}} \\ 
\frac{1}{2} \left( 1 - \lambda \right) &  \frac{1}{2} \left( 1 + \lambda \right) &   \frac{1}{\sqrt{2}} 
\end{array}
\right]  + {\cal O}(\lambda^{2})
\end{equation}
where will use $\lambda$ alternatively to refer to the Cabibbo angle.
We note that the mixing angle $\theta_{13}$ is at present 
constrained to be $\theta_{13}< 0.15 \simeq 3 \lambda^{2} $
by the non-observation of neutrino oscillations at the CHOOZ experiment \cite{Apollonio:2002gd}
and a fit to the global data \cite{Maltoni:2004ei}.   
There will be many efforts in the near future to measure all parameters in the neutrino mixing
matrix to a much higher degree of precision. In the immediate future, reactor neutrino experiments
will strengthen the bounds on $\theta_{13}$ or will actually measure a non-zero value for it.
Is it possible that $\theta_{13}$ is actually zero or should we expect that future experiments
will measure a non-zero value ?  Various symmetry schemes have been proposed in the literature
that lead to a prediction for $\theta_{13}$ while generating a near bimaximal 
MNSP matrix \cite{ue3pred}. 
On the other hand, it is known that the MNSP 
mixing matrix in a general leptonic basis receives contributions 
from both the neutrino and the charged lepton mass matrices,
\begin{equation}
{\cal V}_{\rm MNSP}=
({\cal V}^{l}_{L})^{\dagger} {\cal V}_{\nu},
\label{MNSP}
\end{equation}
where ${\cal V}_{\nu}$ is the neutrino diagonalization matrix
and ${\cal V}^{l}_{L}$ is the left handed charged lepton diagonalization matrix,
${\cal M}_{l}^{\rm diag} = ({\cal V}^{l}_{L})^{\dagger}{\cal  M}_{l} {\cal V}_{R}^{l}$.
It is the main purpose of this paper to show that, irrespective of what is the 
precise nature of the underlying symmetry that determines the exact deviation from
bimaximality in the neutrino mass matrix,
the existence of precise empirical relations between
the charged lepton spectra, the quark spectra and the Cabibbo angle indicates 
that the associated charged lepton mixing together with a 
near bimaximal neutrino mixing matrix must
generate a contribution to $\left|{\cal U}_{e3}\right|$,
It is plausible that this contribution 
is the dominant source of $\left|{\cal U}_{e3}\right|$. 
Here ${\cal U}_{aj}$, ($a=e,\mu,\tau$ and $j=1,2,3$) denote the elements of the MNSP matrix.
\section{A prediction for $\left|{\cal U}_{e3}\right|$}
There is an empirical relation which has been known for quite a long time
\cite{Cabibborelation,Georgi:1979df},
\begin{equation}
\left| V_{us} \right|
\approx
\left[ \frac{m_{d}}{m_{s}} \right] ^{\frac{1}{2}}
\approx 3 \left[ \frac{m_{e}}{m_{\mu}} \right]^{\frac{1}{2}},
\label{pattern1}
\end{equation}
This relation has been recently analyzed with precision by one of the authors
who noted that indeed the relation surprisingly works at the level of $\pm 16$\%, as the following
ratio shows (see Ref.~\cite{javier} for details), 
\begin{eqnarray}
\left[ \frac{m_{d}}{m_{s}}\right] ^{1/2} :
\left[  \frac{m_{e}}{m_{\mu}} \right] ^{1/2}
&=& 3.06 \pm 0.48.
\label{Cabibbolepton} 
\end{eqnarray}
The relation between the Cabibbo angle and the down-strange quark mass ratio
can be simply explained, as known from the '70's\cite{Weinberg:hb},
if the down quark mass is generated from the mixing between
the first and second  families. 
In this case, one expects that there is a quark basis where
the normalized down-type quark mass matrix is given to leading order by,
\begin{equation}
\widehat{\cal M}_{D}=
\left[
\begin{array}{ccc}
0 & \left( \frac{m_{s} m_{d}}{m^{2}_{b}}  \right)^{\frac 1 2 }  &  {\cal O}(\lambda^{3})  \\
   \left( \frac{m_{s} m_{d}}{m^{2}_{b}}  \right)^{\frac 1 2 }    & 
  \left( \frac{m_{s}}{m_{b}} \right)  &  {\cal O}(\lambda^{2})  \\
  {\cal O}(\lambda^{3})   &  {\cal O}(\lambda^{2})  & 1 
\end{array}
\right].
\label{MDmat}
\end{equation}
The order of magnitude in the coefficients $(\widehat{\cal M}_{D})_{13}$ and 
$(\widehat{\cal M}_{D})_{23}$ is obtained by requiring these entries not
to affect the quark mass ratios predicted by the matrix to
leading order.  Analogously, the relation between the Cabibbo angle 
and the electron-muon mass ratio can also be simply explained
if the electron mass is generated from the mixing between
the first and second lepton families. This implies that there is a leptonic
basis where the charged lepton mass matrix is given to leading order by,
\begin{equation}
\widehat{\cal M}_{l}=
\left[
\begin{array}{ccc}
0 & \left( \frac{m_{\mu} m_{e}}{m^{2}_{\tau}}  \right)^{\frac 1 2 }  &  {\cal O}(\lambda^{3})  \\
   \left( \frac{m_{\mu} m_{e}}{m^{2}_{\tau}}  \right)^{\frac 1 2 }    & 
  \left( \frac{m_{\mu}}{m_{\tau}} \right)  &  {\cal O}(\lambda^{2})  \\
  {\cal O}(\lambda^{3})   &  {\cal O}(\lambda^{2})  & 1 
\end{array}
\right].
\label{MLmatrix}
\end{equation}
The order of magnitude in the 
coefficients $(\widehat{\cal M}_{l})_{13}$ and 
$(\widehat{\cal M}_{l})_{23}$ can be obtained by requiring these entries 
to not modify the leading order terms for the charged lepton mass ratios.
Such a form for the charged lepton mass matrix is also obtained in the
mass matrix ansatz in ref.~\cite{Bjorken:2002vt}.
From the matrix in Eq.~\ref{MLmatrix} and the empirical 
relation in Eq.~\ref{pattern1} follows that the 
charged lepton mixing matrix is given in this leptonic basis by, 
\begin{equation}
{\cal V}_{L}^{l}=
\left[
\begin{array}{ccc}
1 & \lambda/3
& {\cal O}(\lambda^{3}) \\
\lambda/3
 & 1 & {\cal O}(\lambda^{2}) \\
{\cal O}(\lambda^{3}) & {\cal O}(\lambda^{2}) & 1 
\end{array}
\right] 
\label{chlepmix}
\end{equation}
It is known that the relation in Eq.~\ref{pattern1} between the quark masses, 
the 
charged lepton masses and the Cabibbo angle 
could be explained in some GUT models
\cite{Georgi:1979df}. In that case it is plausible that the basis where 
${\cal V}_{L}^{l}$ adopts the form given by Eq.~\ref{chlepmix}
while the down-type quark mass matrix adopts the form given by Eq.~\ref{MDmat} 
is the gauge flavor basis of the GUT model where quark and leptons unify in 
the same representations.
Let us assume that the charged lepton mixing matrix
is given to leading order in $\lambda$ by Eq.~\ref{chlepmix}
and in the same basis the underlying neutrino mass matrix generates an
exactly bimaximal neutrino mixing matrix, 
\begin{equation}
{\cal V}_{\nu} =  \left[
\begin{array}{ccc}
 \frac{1}{\sqrt{2}}  & - \frac{1}{\sqrt{2}}& 0 \\
 \frac{1}{2}&  \frac{1}{2} & - \frac{1}{\sqrt{2}} \\ 
\frac{1}{2} &  \frac{1}{2}  &   \frac{1}{\sqrt{2}} 
\end{array}
\right]. 
\end{equation}
In this case one expects the charged lepton mixing to induce a non-zero 
$\left|{\cal U}_{e3}\right|$ \cite{Romanino:2004ww}. In our case,
as long as the mixing in the neutrino sector
is approximately bimaximal, we find, 
\begin{equation}
\left| {\cal U}_{e3}\right| =  \frac{\lambda}{3 \sqrt{2}} = 
\left( \frac{m_{e}}{2 m_{\mu}} \right)^{\frac{1}{2}}
 \approx 0.052 \pm 0.001
\end{equation}
The present fit to the global data indicates that $\sin \theta_{13}<0.15$ 
at $90$\% C.L. \cite{Maltoni:2004ei}.
There are some reactor experiments proposed for the
future: BRAIDWOOD 
in Illinois, DAYA BAY in China and
KASKA in Japan, that are expected to reach the level of $\sin \theta_{13}\approx 0.05$
\cite{wang}.  
Since CHOOZ II will only reach
a sensitivity in $\sin \theta_{13}$ of $\approx 0.08$ at $90$\% C.L. after 3 years of operation \cite{choozII}, 
we expect it to obtain a null result.
It has been estimated that neutrino 
factories will reach values of the order $\left|{\cal U}_{e3} \right|\approx0.025$
\cite{neutrinofactories}.  
This prediction is rather robust as it will follow even if the neutrino mixing
matrix is not bimaximal, but merely if the third column has the form as in 
eq(13). For example, a neutrino mixing matrix of the so-called tri-bimaximal
\cite{harrison} form will also yield the same result.
We have learnt during the elaboration of this paper 
of a simultaneous derivation of this prediction by 
J.D.~Bjorken \cite{bjresult} 
in the context of the model proposed in Ref.~\cite{Bjorken:2002vt} 
\section{A prediction for $\sin \theta_{23}$}
In this section we would like to point out that based on a second
empirical relation between the fermion masses and the CKM elements
recently unveiled \cite{javier,javierhaba} it is plausible to expect
also a contribution to $\sin (\theta_{23})$, coming from the (23)
mixing in the charged lepton mass matrix, and a non-zero CP-violating phase
in the MNSP matrix. Nevertheless, the predictions in this section for 
$\theta_{23}$ and $\delta$ are less robust than the one for $U_{e3}$. 
The new empirical relation mentioned above is given by,
\begin{equation}
\theta \approx
\left[ \frac{m_{s}^{3}}{m_{b}^{2}m_{d}} \right] ^{\frac{1}{2}}
\approx \left[ \frac{m_{c}^{3}}{m_{t}^{2}m_{u}} \right] ^{\frac{1}{2}}
\approx  \frac{1}{9} \left[ \frac{m_{\mu}^{3}}{m_{\tau}^{2}m_{e}} \right]^{\frac{1}{2}}.
\label{pattern2}
\end{equation}
This relation together with an additional empirical 
relation with the quark mixing angles,
\begin{eqnarray}
\theta &\approx & \frac{1}{2} \left|\frac{V_{cb}}{V_{us}}\right| = 0.093 \pm 0.003,
\end{eqnarray}
implies \cite{javier} that the quark mass matrices can be reconstructed to leading order
as a function of the two basic flavor parameters: $\theta$ and $\lambda$.
In certain basis where the up-type quark mass matrix is diagonal 
the reconstructed normalized down-type quark matrix would be given by, 
\begin{equation}
\widehat{\cal M}_{d}=
\left[
\begin{array}{ccc}
0 & \theta \lambda^{2} & \theta \lambda^{2} e^{-i \gamma} \\
 \theta \lambda^{2} & \theta \lambda & 2 \theta \lambda \\
 \theta \lambda^{2} e^{i \gamma}  & 2 \theta \lambda & 1 
\end{array}
\right],
\end{equation}
where $\gamma$ is the standard CP-violating phase.
Here $\widehat{{\cal M}}_{d}$ is bidiagonalized by 
$( {\cal V}^{d}_{L})^{\dagger} \widehat{{\cal M}}_{d} {\cal V}_{R}^{d}$. 
In this quark basis, ${\cal V}_{\rm CKM}={\cal V}_{L}^{d}$, which to leading order is, 
\begin{equation}
{\cal V}_{\rm CKM}=
\left[
\begin{array}{ccc}
1 - \lambda^{2}/2 & \lambda & - \theta \lambda^{2} e^{-i \gamma} \\
-\lambda & 1 - \lambda^{2}/2 &  -2 \theta \lambda \\
( e^{i \gamma} -2 )\theta \lambda^{2}   & 2 \theta \lambda & 1 - 2 \theta^{2} \lambda^{2}
\end{array}
\right] 
\end{equation}
It has been shown \cite{javier}
that this simple mass matrix $\widehat{\cal M}_{d}$ 
fits all the experimental data with precision and additionally predicts
a simple succesfull relation between the quark CP phases, 
$\beta= {\rm Arg} \left[ 2 - e^{- i \gamma} \right]$.
If there is a connection between the charged lepton mass matrix
and the down-type quark matrix, as is the case in some GUT models
\footnote{For instance in SU(5) models where 
the Higgs field giving mass to the charged leptons and down-type quarks
transforms under the representation {\bf 45} of SU(5) \cite{Georgi:1979df}},
we expect that there is a leptonic basis where the normalized
charged lepton mass matrix is given by,
\begin{equation}
\widehat{\cal M}_{l}=
\left[
\begin{array}{ccc}
0 & \theta \lambda^{2} & \theta \lambda^{2} e^{-i \gamma} \\
 \theta \lambda^{2} & 3 \theta \lambda & 2 \theta \lambda \\
 \theta \lambda^{2} e^{i \gamma}  & 2 \theta \lambda & 1 
\end{array}
\right].
\end{equation}
In this basis the charged lepton mixing matrix would be given
to leading order by,
\begin{equation}
{\cal V}_{L}^{l}=
\left[
\begin{array}{ccc}
1  & \lambda/3 & - \theta \lambda^{2} e^{-i \gamma} \\
-\lambda/3 & 1  &  -2 \theta \lambda \\
(e^{i \gamma} - \frac 2 3 ) \theta \lambda^{2} 
  & 2 \theta \lambda & 1 
\end{array}
\right], 
\label{Vlcomp}
\end{equation}
where the deviation from unitarity is at most of order $\theta \lambda^{3}$.
If we assume that the neutrino mixing matrix is exactly bimaximal (or rather
the third column has that form), we obtain, using Eq.~\ref{MNSP}, a prediction 
for $\sin ^{2} (2\theta_{23})$ given by,
\begin{equation}
\sin ^{2} (2\theta_{23}) = 4 \left| {\cal U}_{\mu 3} {\cal U}_{\tau 3}\right|^{2}  
= 1 - \frac{4}{9}\left( \frac{m_{\mu}}{m_{\tau}}\right)^{2}.
\label{theta23dev}
\end{equation}
This corresponds to $\sin ^{2} (2\theta_{23}) \approx 0.998$,
or that $\theta_{23}$ differs
from $\pi/4$ by $\simeq 1.7^{\circ}$.
We expect that future experiments could rule out this prediction
for $\theta_{23}$.
The magnitude of the CP-violating effects in neutrino oscillations
is controled by the rephasing invariant $J_{CP}^{\nu}= 
{\rm Im} \left[ {\cal U}^{*}_{km} {\cal U}_{lm}
{\cal U}_{kn} {\cal U}^{*}_{ln}\right]$ (irrespective of the indices).
If the neutrino mixing matrix was nearly bimaximal and CP conserving 
the source of the CP-violating phase in the MNSP matrix would 
arise from the phase present in the charged lepton mixing matrix 
\footnote{The observation that a hierarchical structure in the charged
lepton mixing matrix would supress a possible contribution to $J^{\nu}_{CP}$ 
has been simultaneously made by Ref.\cite{Petcov:2004rk}},
which based on the matrix in Eq.~\ref{Vlcomp} is given by,
\begin{equation}
J^{\nu}_{CP} = - \frac{\theta \lambda^{2}}{4 \sqrt{2}} \sin \gamma.
\label{JCPL}
\end{equation}
The phase $\delta$ would be given in this case 
by $\tan \delta = 3 \lambda \theta \sin \gamma$.
Experimentally $\gamma$ seems to be a large angle \cite{Eidelman:2004wy}.
The 2004 winter global fit of the CKM elements obtained 
using the program CKMFitter \cite{Hocker:2001xe} gives us
$\gamma_{\rm exp}  =  61^{\circ} \pm 11 ^{\circ}$. Therefore
$J_{CP}^{\nu}$ could be as large as about $10^{-3}$,
which corresponds to a phase $\delta \approx 3^{\circ}$.
Nevertheless, the CP-conservation of the neutrino mixing matrix
is a very strong assumption. It is known that if neutrinos are Majorana particles 
some neutrino phases cannot be absorbed by redefinition
of the neutrino fields \cite{Majophase} 
and in general there would be a contribution to $J^{\nu}_{CP}$ given by,
\begin{equation}
J^{\nu}_{CP} = - \frac{\lambda \sin \phi}{12 \sqrt{2}}.
\label{JCP}
\end{equation}
where $\phi$ is one of the Majorana phases. 
If $\sin \phi$ is near one, this contribution would be dominant over the
one in Eq.~\ref{JCPL} and would give a maximum $J_{CP}^{\nu}$ 
as large as $0.014$, which corresponds to $\delta \approx 45^{\circ}$. 
We expect that, irrespective of the nature of the neutrino, 
future experiments have to measure a value of $\delta$
between the two limits given by Eq.~\ref{JCPL} and Eq.~\ref{JCP},
{\it i.e.}  $ 3^{\circ} < \delta < 45^{\circ}$.
\section{Can the QLC relation arise from charged lepton mixing ?}
The presence of the Cabibbo angle in the MNSP matrix,
as recent measurements of the solar mixing angle indicates,
at first sight may suggest  that all  deviations from the 
exact bimaximal ansatz may be a contamination coming from 
the charged lepton mixing matrix. 
We have seen that the patterns in the fermion spectra suggest that
there is a leptonic basis where the electron mass is generated from
the mixing between the first two flavor families. This basis is most probably
the gauge flavor basis of the a theory where quarks and leptons 
unify in common representations. It is precisely in this basis where
one would expect the neutrino mixing matrix to be exactly bimaximal.
Nevertheless, if this was the case we would obtain that 
$\theta_{12} = \frac{\pi}{4} + \frac{\theta_{C}}{6}$ 
instead of the experimentally observed 
$\theta_{12} = \frac{\pi}{4} - \theta_{C}$,
too small and of the opposite sign required to account for the QLC relation. 
If one insists to fully generate the observed deviation from 
bimaximality in the MNSP matrix from the charged lepton mixing, 
the required mixing would be very large and as a consequence in such
a basis the charged lepton mass matrix would adopt a very unnatural form
in order to reproduce the correct electron mass \cite{Frampton:2004ud}.
Therefore, we believe  that most probably the 
Cabibbo angle is already present in the neutrino mass matrix,
or in other words the QLC relation must arise from the mechanism 
that generates the neutrino mass matrix and not
from the charged lepton mixing.
Of course, it is entirely possible that the QLC relation is
only approximate and furthermore is accidental and a red herring and
does not therefore need any explanation.
\acknowledgements
We especially thank J.D.~Bjorken and also 
L.J.~Hall and W.~Rodejohann for comments and suggestions.
This work is supported by:
the Director, Office of Science, Office of High Energy and Nuclear Physics, 
of the US Department of Energy under Contracts DE-AC03-76SF00098, 
DE-FG03-91ER-40676 and DE-FG03-94ER40833,
by the National Science Foundation under grant PHY-0098840 and by 
the Ministry of Science of Spain under grant EX2004-0238.

\end{document}